\documentclass[a4paper,english]{revtex4}
\usepackage[T1]{fontenc}
\usepackage[latin9]{inputenc}
\usepackage{color}
\usepackage{amsmath}
\usepackage{amssymb}
\usepackage{wasysym}
\usepackage{graphicx}
\usepackage{esint}
\PassOptionsToPackage{normalem}{ulem}
\usepackage{ulem}

\makeatletter

\pdfpageheight\paperheight
\pdfpagewidth\paperwidth

\@ifundefined{textcolor}{}
{%
 \definecolor{BLACK}{gray}{0}
 \definecolor{WHITE}{gray}{1}
 \definecolor{RED}{rgb}{1,0,0}
 \definecolor{GREEN}{rgb}{0,1,0}
 \definecolor{BLUE}{rgb}{0,0,1}
 \definecolor{CYAN}{cmyk}{1,0,0,0}
 \definecolor{MAGENTA}{cmyk}{0,1,0,0}
 \definecolor{YELLOW}{cmyk}{0,0,1,0}
}

\@ifundefined{definecolor}
 {\usepackage{color}}{}
\@ifundefined{definecolor}{\usepackage{color}}{}
\usepackage{babel}

\newcommand{\NOTE}[1]{\textbf{\textcolor{red}{}}}
\newcommand{\ADD}[1]{{#1}}
\newcommand{\MOD}[1]{{#1}}
\newcommand{\REM}[1]{\textcolor{red}{}}
\newcommand{\etal}{\mbox{\textit{et al. }}}

\usepackage{babel}

\usepackage{babel}

\makeatother

\usepackage{babel}
\begin{document}

\title{Caustics and clustering in the vicinity of a vortex}

\author{S. Ravichandran and Rama Govindarajan}

\affiliation{TIFR Centre for Interdisciplinary Sciences \\
 Tata Institute of Fundamental Research, Narsingi, Hyderabad, 500075,
India. \\
 ravis@tifrh.res.in, rama@tifrh.res.in. }
\begin{abstract}
We study the formation of caustics in vortex-dominated flows. We find
that only particles starting within a critical distance of a vortex
which scales as the square roots of the particle inertia and the circulation
can form sling caustics. We show that particles starting in an annular
region around this critical radius contribute the densest clusters
in the flow. The large density spikes occurring for such particles,
even at small inertia, are indicative that these particles will experience
large collision rates. 
\end{abstract}
\maketitle

\section{Introduction\label{sec:Introduction}}

When a turbulent flow contains small particles of a different density
dispersed in it, these particles tend to cluster into small regions
of the flow rather than being homogeneously distributed in space \cite{Bec2003,Bec_etal_2007,Bec_etal_2005,Eaton_Fessler,Shaw1998}.
Several different but related ways of explaining this clustering have
been posited \cite{ChenGotoVassilicos2006,GibertXu2012,Goto2008,Sapsis&Haller2010_Chaos}.
Typically, heavy particles are centrifuged out of vortical regions
and cluster into regions of high strain \ADD{\cite{Maxey_1987}}.
\ADD{This effect can observed for particles that are larger (typical
of laboratory studies, see e.g. \cite{Eaton_Fessler}) or smaller than the Kolmogorov scale. In particular,
the behaviour of heavy particles much smaller than the Kolmogorov
scale can be described using a simple equation (see eq. (1)) and is
important in atmospheric flows.} \REM{The fact that heavy particles
cluster is of importance to many processes in atmospheric and oceanographic
flows, e.g. in rain formation and plankton population dynamics, has
motivated several studies to understand this process better.}

\REM{We are interested in one particular feature of such a flow:
the formation of caustics due to centrifugal forces near vortices.}
\ADD{Particle inertia also leads to the formation of caustics.}
At a given time instant, the velocities of two neighboring particles
can differ greatly from each other. Velocity contours of inertial
particles may thus display discontinuities and multi-valuedness in
some regions of the flow. Such regions are known as caustics, where
particle velocity may not be described as a \ADD{single-valued}
field. The effects of caustics have been extensively studied \cite{M-W 2005EPL,M-W 2005PRL,M-W 2006 PRL,M-W 2007 PoF,Mehlig 2010 EPL,Mehlig 2011 EPL,Mehlig 2012 NJP,Mehlig 2013 PRE,Falkovich 2007 JAS}.
For example Mehlig and Wilkinson \cite{M-W 2005EPL} describe \ADD{caustics
as occuring due to} \REM{caustics which occur because of} folds
in the velocity in phase space, leading to points in space where there
are multiple values for the velocity at the same instant of time.
Falkovich \etal \cite{Falkovich Nature} describe \ADD{the
same phenomenon} as the formation of caustics due to the `sling effect'
or centrifugal action near the vortices. Caustics increase the collision
rates for particles drastically, and \REM{are thus thought to be
primarily responsible for} \ADD{have been suggested as a mechanism
for} the suddenness of the onset of rain in convecting cumulus clouds
\cite{M-W 2005EPL,M-W 2006 PRL,Falkovich Nature}.

\ADD{We ask how important the processes of sling caustics and preferential
concentration are relative to each other.} \REM{We ask whether there
are conditions under which sling caustics can form and conditions
under which they will not.} To this end we examine the archetypal
flow for the sling effect: the flow due to a point vortex. In the
process of being centrifuged out of vortical regions, particles which
are closer to a vortex attain higher acceleration, and can eventually
catch up with, or even overtake, particles which were initially further
away from the vortex, thus forming caustics. \REM{We examine here
how important these caustics are to the clustering process. Mehlig
\& Wilkinson and others have argued using numerical simulations of
random flows that caustics are responsible for a large fraction of
the clustering of particles for Stokes number $St\sim1$, and that
the centrifugal effect of particles being flung out of vortical regions
is only important for $St\ll1$.} We show that caustic-formation
is only possible for particles initially located within a critical
distance $r_{cr}\simeq1/2\sqrt{\Gamma\tau}$ of a point vortex, where
$\Gamma$ is the circulation of the vortex and $\tau$ is the relaxation
time of the particle\ADD{s, such that the particle Stokes number 
at the critical radius, $St = \Gamma\tau/r_{cr}^2$, is of order $1$.} Interestingly we find that
the largest contributions to clustering come from the band of particles
located in an annular region around this critical radius. This has
consequences both for collision of particles and for condensation
inhomogeneities.

We note that the system of particles around a vortex has been studied
previously by Raju \& Meiburg\cite{Raju&Meiburg}. They provide a
detailed discussion of the different particle-inertia regimes involved,
and find similar spikes in density as we report in section \ref{sec:three}.
They also report an increased ``accumulation rate'' for intermediate
values of the particle Stokes number. Their work was done before the
idea of inertial particle caustics came about, and we find that caustics
provide a way of understanding their results.

In section \ref{sec:two}, we discuss the general nature of the problem,
including the different possible initial conditions that can be used.
In section \ref{sec:three}, we look at point vortices in particular.
We discuss when caustics can form in the flow around a vortex, and
show some scaling laws that are obeyed. Since point vortices are idealisations
and viscosity smears them out, we examine caustics formation due to
Gaussian vortices in \ref{sec:four}. In section \ref{sec:five},
we report simulations with many vortices in a doubly-periodic box.
We conclude in section \ref{sec:six}.

\section{Caustics around a vortex}

\label{sec:two}

We prescribe a two-dimensional steady axisymmetric flow field, with
its centre at the origin of the coordinate system. The radius $a$
of the particle is taken to be much smaller than the typical flow
length scales. The motion of such an inertial particle in a flow field
$\mathbf{u}\left(\mathbf{x},t\right)$ is given by the Maxey-Riley
equations \cite{Maxey-Riley}\ADD{. For particles much smaller than the Kolmogorov scale, 
and much denser than the fluid, the Maxey-Riley equations simplify to} 
\begin{align}
\frac{d\mathbf{r}}{dt} & =\mathbf{v}\nonumber \\
\frac{d\mathbf{v}}{dt} & =\frac{\mathbf{u}-\mathbf{v}}{\tau},\label{eq:inertial_general}
\end{align}
 where $\mathbf{r}$ and $\mathbf{v}$ are the particle location and
velocity respectively, and $t$ is time. \ADD{The effect of particle
inertia is encapsulated in the particle time-lag $\tau$.} \REM{In
the heavy particle limit we have a single time scale of relaxation
$\tau$, given by the Stokes drag on the particle.}

The most general axisymmetric flow field is given by 
\begin{equation}
\Omega\sim\frac{1}{r^{p}}\quad{\rm or}\quad\left|\mathbf{u}\right|=r\Omega\sim r^{1-p},\label{eq:general_vortex_flow}
\end{equation}
 where $\Omega\left(r\right)$ is the angular velocity of a fluid
particle at a radius $r$. For solid body rotation, $p=0$, whereas
for a flow with a single point vortex placed at the origin, $p=2$.
In cylindrical polar ($r,\theta$) coordinates equation (\ref{eq:inertial_general})
is written, with 
\begin{equation}
\omega\equiv d\theta/dt\ ,\label{eq:omega_defn}
\end{equation}
 as 
\begin{eqnarray}
\dot{r} & = & v_{r}\nonumber \\
\dot{v}_{r}+\frac{v_{r}}{\tau} & = & r\omega^{2}\nonumber \\
\dot{\left(r^{2}\omega\right)}+\frac{r^{2}\omega}{\tau} & = & \frac{r^{2}\Omega}{\tau}.\label{eq:inertial particle general vortex}
\end{eqnarray}
 where the dot superscript stands for $d(.)/dt$. Particle dynamics
in the vicinity of one vortex has been studied using the above equations
in some detail by Raju \& Meiburg \cite{Raju&Meiburg} and by Shaw
\etal \cite{Shaw1998}. \REM{The latter
show how droplet distributions are affected by the centrifuging due
to the vortex. They also have an excellent discussion about
the different time scales in the problem.}

From these solutions it is known that the radial distance of a particle
from the center of the coordinate system is a monotonically increasing
function of time. At long times the tangential velocity of a particle
will \REM{asymptote} \ADD{approach} that of the fluid, i.e., $\omega\simeq\Omega$
and so from equation (\ref{eq:inertial particle general vortex}),
we have at long time or large radial distance, $v_{r}\simeq\tau r\Omega^{2}$
and \REM{particle} \ADD{the particle's radial} acceleration is
negligible.

We prescribe two rings of particles at initial radii $r_{01}$ and
$r_{02}$ with initial angular velocities $\omega_{i}=\omega_{0}(r_{i})\mbox{ , }i=1,2$.
Equations (\ref{eq:inertial particle general vortex}) then give 
\begin{align}
\frac{d\Delta r}{dt} & =\Delta v_{r}\nonumber \\
\frac{d\Delta v_{r}}{dt}+\frac{\Delta v_{r}}{\tau} & =\Delta\left(r\omega^{2}\right)\nonumber \\
\frac{d}{dt}\Delta\left(r^{2}\omega\right)+\frac{\Delta\left(r^{2}\omega\right)}{\tau} & =\frac{\Delta\left(r^{2}\Omega\right)}{\tau}\label{eq: analysis general}
\end{align}
 where $\Delta(.)=(.)_{2}-(.)_{1}$ represents a difference in a given
quantity between the two rings of particles. In the following we define
$\Delta r_{0}\equiv\Delta r(t=0)$ and use the shorthand notation
$r_{0}$ for the initial location $r_{01}$ of the inner ring.

\begin{figure}
\includegraphics[clip,scale=0.4]{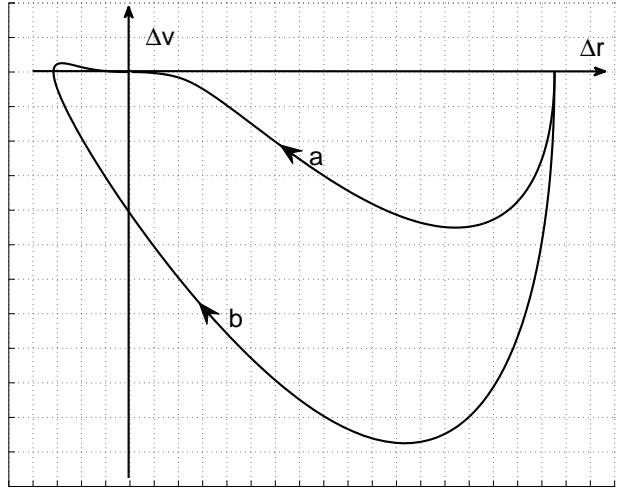}
\caption{\label{fig:Possible-trajectories}\ADD{Schematic of} possible trajectories
for the system (\ref{eq: analysis general}). At long times, the two
rings of particles will come closer to each other and merge at infinite
time, and this would, in the absence of caustics, give rise to a trajectory
of type (a). In the caustics forming case, the ring closer to the
origin at initial time would catch up with and overtake the ring further
out, so the relative velocity would cross zero once in finite time,
as in trajectory (b).}
\end{figure}

A caustic can be said to have formed if, at a particular finite time
$t$, particles on both rings arrive the same radius $r$ but with
different velocities, i.e. if $\Delta r=0$, but $\Delta v_{r}\ne0$.
Note in particular that particles which asymptotically approach each
other at long times \REM{will} \ADD{do} not give rise to caustics,
although they \REM{will} \ADD{do} contribute to clustering. When
a caustic occurs, trajectories in the phase portrait for the relative
position and velocity \REM{will} look like the curve `b' in figure
\ref{fig:Possible-trajectories} and not like curve `a'.

We can solve the system (\ref{eq: analysis general}) for the general
vortex $\Omega\sim r^{-p}$ and \REM{demarcate the boundary} \ADD{plot
the separatrix} in $r_{0}-\Delta r$ space between caustic-forming
and no-caustic regimes. We only show results for $p=2$, and comment
on the implications of $p\neq2$ (see end of section \ref{sec:four}).

We use three kinds of initial conditions for this study. The first,
one we shall call co-rotating (or `zero-inertia'), that the particles
follow the flow exactly until $t=0$, and particle inertia is switched
on at $t=0$. For particles of small inertia this is a reasonable
condition \cite{Raju&Meiburg}. \REM{Also in a cloud, there are aerosol
particles which are, to a very good approximation, merely tracer particles
which follow the flow. Once water begins to condense on to these particles,
their inertia increases. The simplest approximation is to switch on
a constant inertia instantaneously.} In the second initial condition,
particles are initially motionless and the vortex is introduced at
$t=0$. We shall refer to this as the zero-velocity initial condition.
A turbulent patch moving into a particle-rich region would correspond
to this situation. Most studies of particles in turbulence employ
one of these two conditions. Also consider a vortex tube being stretched
by the turbulent flow around it. The strength of the vortex will grow
with time until it reaches it\ADD{s} maximum value. To mimic this
situation, a third initial condition, which is just a generalisation
of the second, is where both particles and vortex are initially motionless,
and the strength of the vortex is gradually ramped up to its final
value. The results from this third initial condition are not presented
here because they display no unexpected physics. The introduction
of another time scale of growing vorticity merely scales the answers
appropriately. Most real-life situations will correspond with reasonable
fidelity to one of our initial conditions or something in-between,
so we expect our results to have a bearing on understanding clustering.

The time and length scales inherent to the problem respectively are
the inertial timescale $\tau$ of the particle and $L=\sqrt{\Gamma\tau}$
where $\Gamma$ is the strength of the vortex. \ADD{Note that $\Gamma$ and $\tau$ are the only parameters in the problem.} For a given initial condition, from dimensional arguments we may argue that the caustic-formation time $T_{caustic}$ will \ADD{have to depend on nondimensional groups formed out of the scales of the problem}, and must have a functional form
\begin{equation}
\frac{T_{caustic}}{\tau}=T\left(\frac{r_{0}}{\sqrt{\Gamma\tau}}\mbox{, }\frac{\Delta r}{\sqrt{\Gamma\tau}}\right).\label{eq:caustic time scaling}
\end{equation}
 If there is indeed a regime where it is impossible for caustics to
form, the caustic-formation time must diverge in this regime. We may
then define a critical initial separation $\Delta r_{cr}$, as the
maximum value of the initial separation for a given $r_{0}$ up to
which caustics can occur. This quantity, suitably scaled, will be
a function only of the initial distance from the centre of the vortex,
i.e., 
\begin{equation}
\frac{\Delta r_{cr}}{\sqrt{\Gamma\tau}}=f\left(\frac{r_{0}}{\sqrt{\Gamma\tau}}\right).\label{eq:scaling of delr}
\end{equation}

We study in detail the case of a single point vortex, and then examine
the case of Gaussian vortices. Finally we will derive a necessary
condition for caustics formation for a general axisymmetric flow.

\section{Point vortex}

\label{sec:three}

For a point vortex, we substitute $p=2$ in equation (\ref{eq:general_vortex_flow}).
The flow field is now given by 
\[
\mathbf{u}=\frac{\Gamma}{2\pi r}\mathbf{e}_{\theta}.
\]
 Using $\tau$ and $\sqrt{\Gamma\tau}$ as scales, equations \ref{eq:inertial particle general vortex}
may be written in non-dimensional form as 
\begin{eqnarray}
{\ddot{x}}+{\dot{x}} & = & \frac{\zeta^{2}}{x^{3}}\label{eq:rdd}\\
{\rm and}\qquad{\dot{\zeta}}+\zeta & = & 1,\label{eq:nondim}
\end{eqnarray}
 where $x=r/(\tau\Gamma)^{1/2}$ and $\zeta\equiv r^{2}\omega/\Gamma$.
We note that these equations are independent of $\tau$, i.e., the
existence of caustics can be investigated with the particle size scaled
out. From equation (\ref{eq:nondim}) it is clear that the azimuthal
motion merely relaxes from any initial condition to $\zeta=1$ on
a time-scale of $\tau$. At longer times, and when $x\gg1$, irrespective
of the initial condition, equation (\ref{eq:rdd}) thus reduces to
one where $\ddot{x}$ is negligible and $\dot{x}\sim f(x)$. In such
a regime, given that $v_{r}>0$ (because of the centrifugal nature
of particle motion), it is obvious that two particles which are at
different radii at a given time $t$ in this regime will never produce
a caustic. It follows that only particles which start out at small
radial distance will accelerate enough to form caustics.

\begin{figure}
\includegraphics[width=0.45\paperwidth]{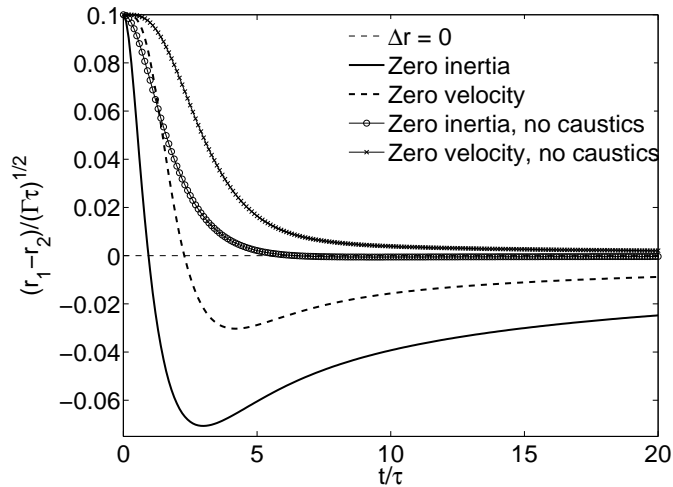}
\caption{\label{fig:initial conditions}The separation between particle-pairs
as a function of time, illustrating the difference between the types
of initial conditions. The curves that do or do not cut the $\Delta r=0$
axis are respectively for particles that do or do not \REM{undergo}
\ADD{form} caustics. Note that a maximum of one crossing of the
$\Delta r=0$ line can occur for a given curve.}
\end{figure}

A numerical solution of equation (\ref{eq:rdd}) enables us to divide
the parameter space into regimes where caustics formation is possible
and regimes where it is not. Figure \ref{fig:initial conditions}
shows examples of both of these, and how the separation between the
two rings of particles progresses with time. 

\subsection{Limits for caustics formation \label{sub:Scaling-of-delr_vs_r0}}

\begin{figure}
\includegraphics[width=0.5\columnwidth]{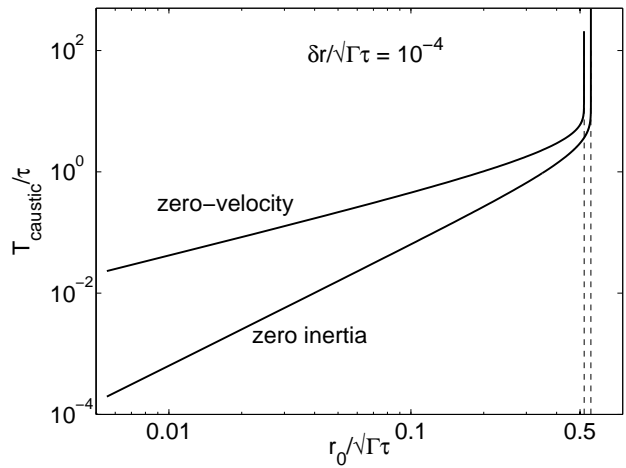}
\caption{\label{fig:blowup of T_collision vs r0}The caustic-formation time
$T_{caustic}$ as a function of the initial location of the inner
ring of particles. A very small separation between the rings is used
in this case, and the location of divergence is insensitive to further
reduction in separation.}
\end{figure}

The time at which each curve crosses the $\Delta r=0$ line is $T_{caustic}$.
We expect the caustic-formation time to be small for small initial
separations of the two rings of particles, and plot this quantity
for $\Delta r_{0}=10^{-4}$ in figure \ref{fig:blowup of T_collision vs r0}.
It is immediately clear that the caustic-formation time diverges even
for this small separation above an \REM{$r_{0}$} \ADD{$x_{0}$}
of about $0.5$, for both initial conditions. We define the initial
radius of the inner ring, above which caustics cannot form, as $r_{cr}$.
In the region where $r_{0}<r_{cr}$, we see that $T_{caustic}$ scales
as $r_{0}^{2}$ for the zero inertia initial condition, and as $r_{0}$ for the zero velocity initial condition.
\ADD{Note that, for the zero inertia initial condition, $T_{caustic}$ is seemingly independent of $\tau$, since $T_{caustic}/\tau$ scales as $(r_{0})^2/(\Gamma\tau)$. After Raju \& Meiburg\cite{Raju&Meiburg}, we can argue for this condition that the initial velocity
term in equation \ref{eq:rdd} is negligible. This gives an equation
for the evolution at early times of the separation between particles that goes as
(assuming $\Delta r\ll r_{0}$):}
\[
\ADD{\frac{d^{2}}{dt^{2}}\Delta r\sim\frac{\Delta r}{r_{0}^{4}}}.
\]
\ADD{This suggests that $T_{caustics}$ should scale as $r_{0}^{2}$ for a given $\Delta r$. For
the zero velocity initial condition, this simplification is not possible,
and we can think of no simple argument at this point for the
scaling as $St^{-1/2}$. }

\begin{figure}
\includegraphics[width=0.5\columnwidth]{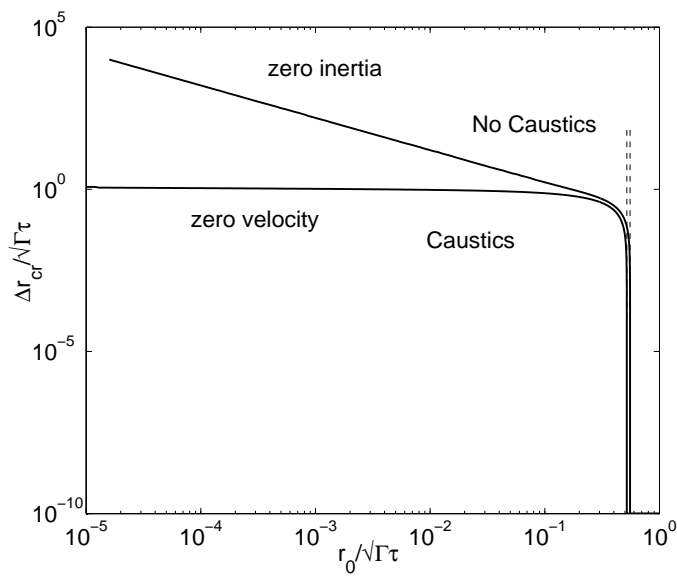}
\caption{\label{fig:Caustics_combined}Maximum initial separation for caustics
formation for the zero-inertia and the zero-velocity initial conditions.
Caustics can occur below of to the left of the lines. The curves for
different $\tau$ collapse onto each other when they are scaled according
to equation \ref{eq:scaling of delr}. For the zero-inertia initial
condition, $r_{0}$ and $\Delta r_{cr}$ are inversely related. }
\end{figure}

Figure \ref{fig:Caustics_combined} shows the maximum initial separation
$(\Delta r)_{cr}$ for caustics to occur, as a function of $r_{0}$.
It is evident that beyond a certain $r_{0}=r_{cr}$, however small
the initial separation, caustics cannot occur. For initial particle
locations within a critical radius, one may have caustics, created
with particles which are located within a certain critical separation
distance $(\Delta r)_{cr}$. This quantity depends on the initial
condition. For a range of $r_{0}$, with initially co-rotating particles,
the initial radius and the maximum separation for caustics obey an
inverse relationship, i.e. $r_{0}(\Delta r)_{cr}/(\Gamma\tau)$ is
constant. The zero-velocity initial condition, on the other hand,
yields a much lower $\Delta r_{cr}$ which varies much more slowly
with $r_{0}$.

\ADD{If the inverse relationship of $(\Delta r)_{cr}$ and $r_{0}$ 
is combined with the scaling argument for $T_{caustic}$, we can obtain a scaling for a "critical" velocity
$(\Delta r)_{cr}/T_{caustic} \sim \Gamma^2\tau/r_{0}^3$. Comparing this with equation (\ref{eq:inertial particle general vortex}) suggests that the particles that bridge the largest gaps move outwards almost ballistically and the acceleration term is unimportant for such particles.}

\subsection{Particle density}

We know that inertial particles in a vortical flow cluster into regions
of high density. We would like to understand the role of sling caustics
in this process. We track the density of inertial particles using
the Lagrangian particle tracking method of Osiptsov following Healy
\& Young \cite{Healy-Young 2005}. This approach has been used in
the past (see \cite{Ijermans_etal_2009,Meneguz_etal_2009}). The
equations are given in the appendix. By this approach, one may track
the density in the vicinity of a given particle while moving with
the particle. The method relies on keeping track of the net divergence
of particle velocities in this moving vicinity. The Osiptsov method
allows for obtaining local densities without actually tracking many
rings of particles\REM{, and we have checked that it is far more
accurate than the results obtained by tracking even an extremely large
number of particles}. \MOD{Figure \ref{fig:Density} (a)} \REM{on the left} shows
density profiles as functions of time in the vicinity of particles
started at different $r_{0}$. Densities in the vicinity of particles
whose initial locations lie within $r_{cr}$, i.e., those which can
participate in caustics formation, are shown by dashed lines, whereas
for particles starting out at $r_{0}>r_{cr}$, i.e., in regions which
do not allow caustics, local densities are shown by solid lines. The
difference between the two is stark. Outside the regime that allows
caustics, all that happens is that particles cluster slowly with time,
and the (Lagrangian) density monotonically increases. On the other
hand, densities of particles starting from within the regime of caustics
formation show a sharp spike around $t=T_{caustic}$ for that $r_{0}$,
and subsequently the Lagrangian density decreases monotonically, since
particles which participate in caustics formation later display trajectories
that diverge in radius. The spikes happen by virtue of a patch of
particles that start out within a small neighbourhood of each other
clumping together. The most striking result in this figure is that
the spike for $r_{0}\sim r_{cr}$ is significantly taller than all
the others, and the clustering remains higher for this case even at
later times. We shall present further evidence below to show that
the annular region close to the critical radius is the most important
for clustering.

\REM{On the right of }\ADD{In figure \ref{fig:Density} (b)}, we show density profiles
as a function of radius at a time of $t=15\tau$ and $t=20\tau$.
By this time, caustics have formed and the participating particles
have since dispersed. In this and following figures, when we allude
to \textquotedbl{}only caustics\textquotedbl{} or \textquotedbl{}excluding
caustics\textquotedbl{}, we refer to initial particle locations $r_{o}<0.8r_{cr}$
and $r_{o}>1.2r_{cr}$ respectively. The values $0.8$ and $1.2$
have been chosen arbitrarily, and the answers do not change qualitatively
when we choose other limits. We refer to initial locations within
$0.8r_{cr}\le r_{o}<1.2r_{cr}$ as the annular region, which will
be shown to contribute significantly to the densest regions. To identify
where particles originated from, different symbols are given in this
figure to the contributions of each of these. We see that there is
a minimum radius below which there are no particles, i.e., the density
is zero. This minimum radius increases slowly with time, at the rate
of $1/x^{3}$. We now discuss the large $r$ portion of this figure.
It is interesting to note that at a given radial location, two possible
values of density are seen. This is because every radial location
contains one set of particles which started at $r_{i}<r_{cr}$ and
another which started at $r_{o}>r_{cr}$. Particles that start close
to the vortex are thrown out so violently that they reach locations
far away from the vortex. Their radial locations are now distributed
over a much large radius. On the other hand, particles which started
at $r_{o}\gg r_{cr}$ have moved much less and display a local density
only slightly different from its initial value. Thus the neighbourhoods
of caustics and non-caustics particles are of low and high density
respectively. The picture changes at lower $r$. Here, the densest
neighbourhoods are of the parcels which started out in the annular
region near $r_{cr}$.

\begin{flushleft}
\begin{figure}
\includegraphics[width=1.0\columnwidth]{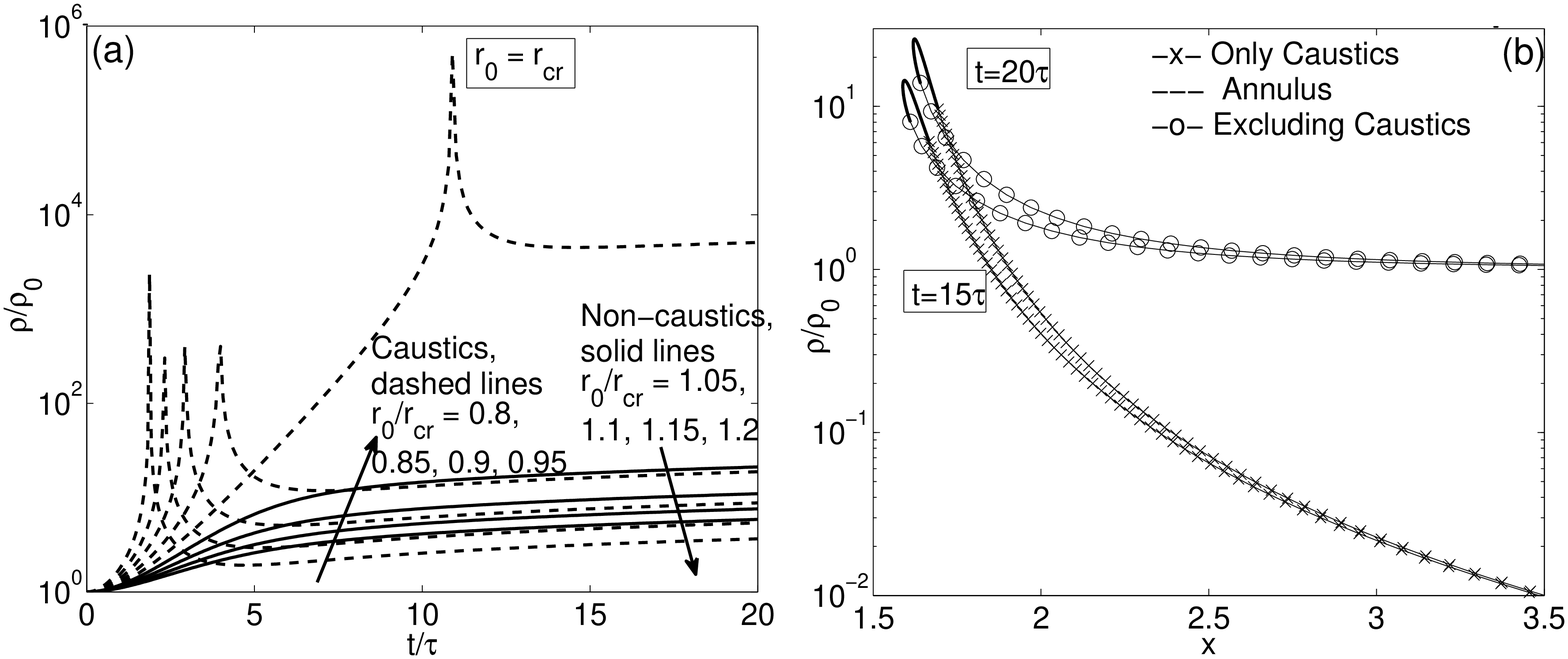}
\caption{\label{fig:Density}Lagrangian densities for initially co-rotating
particles around a single point vortex at the origin. The densities
were tracked using the method of Osiptsov (eq. \ref{eq:Osiptsov_density_tracking}).
\MOD{(a)}: Density as a function of time for different initial location
$r_{0}$. The caustics-producing initial conditions show sharp spikes.
\MOD{(b)}: Particle density as a function of radial distance from the
vortex at $t=15\tau$ and $t=20\tau$. There are two possible densities
at each radial location, one in the vicinity of particles which started
within the caustics-producing region and the other in the neighborhood
of particles which started outside this region.}
\end{figure}

\par\end{flushleft}

\section{Gaussian vortices}

\label{sec:four}

\begin{figure}
\includegraphics[width=1.0\columnwidth]{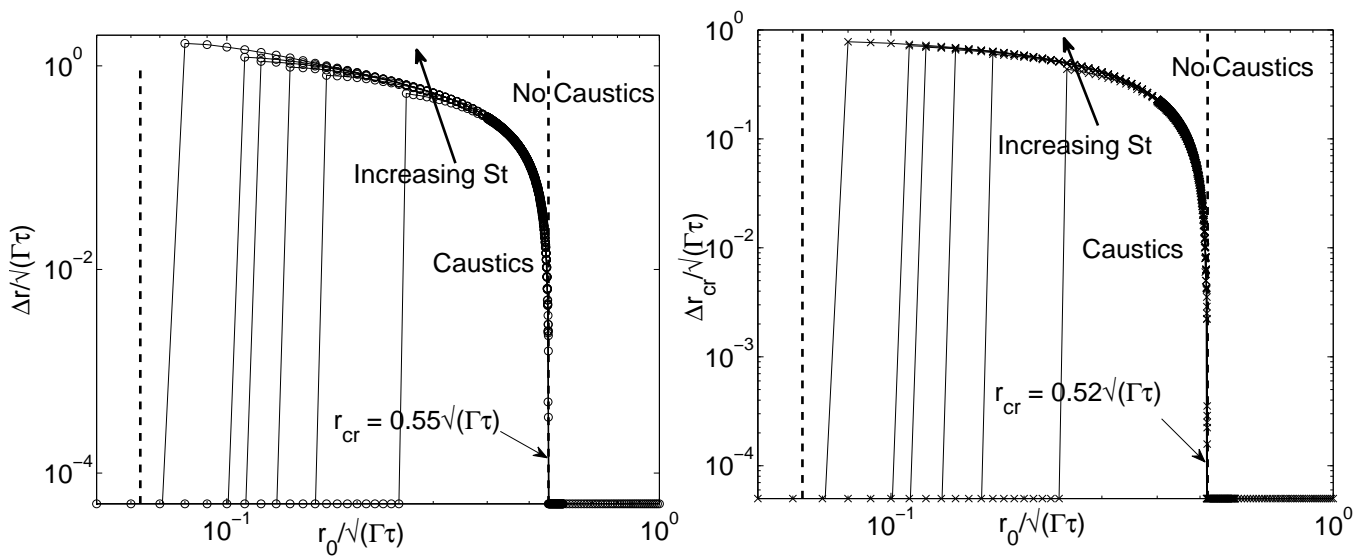}
\caption{\label{fig:Caustics-Gaussian}Caustics with Gaussian vortices. These
plots are analogous to the plots in figure \ref{fig:Caustics_combined}.
The maximum vorticity and characteristic radius are the same for all
plotted curves. The dashed lines to the left of the curves show the
characteristic radius of the Gaussian vortex normalised by the largest
inertia. Caustics exist only for $r_{cr}>r_{0}>1.2\ r_{v}$, and the
inner limit is dependent on $St$ (see text for a definition of $St$).
Left: zero-inertia initial condition, right: zero-velocity. The values
of $St$ shown are 0.04, 0.08, 0.12, 0.16, 0.2 and 0.4.}
\end{figure}

\ADD{Viscosity acts on point vortices by smearing them out into Gaussian
vortices.} \REM{We now look at one possible effect of viscosity
on our results from sections 2 and 3.} \ADD{We therefore study how
caustics form around Gaussian vortices.} We do this by working with
Gaussian vortices of fixed maximum vorticities and widths, such that
their total strength is the same as for the point vortices. The simulations
presented here are inviscid. Figures \ref{fig:Caustics-Gaussian}
plot the caustics boundary for the zero inertia and zero velocity
initial conditions. We find that caustics are still created by particles
initially within the region given by the same critical radius as with
point vortices. However, particles which start out very close to the
centre of the vortex do not participate in caustics formation in this
case, so there is a minimum initial separation of vortex centre and
particle for caustics formation. This radius is dependent on the Stokes
number of the particles. This is because the effective value of the
exponent $p$ in eq. \ref{eq:general_vortex_flow} varies with radius
for a Gaussian vortex. Notice that the independence from particle
inertia $\tau$ of equations (\ref{eq:rdd}) and (\ref{eq:nondim})
is only valid for point vortices, i.e., when $p=2$. For $p$ values
which are lower, we have made computations which show that the results
are Stokes number dependent, and below a particular $p$ for a given
$\tau$, caustics cannot form. For any $\tau$, caustics form only
over a range of $r_{0}$ between $1.2\ r_{v}$ and $r_{cr}$, where
$r_{v}$ is the characteristic radius of a Gaussian vortex, defined
by 
\begin{equation}
\omega(r)=\omega_{v}e^{-r^{2}/r_{v}^{2}},\label{eq:gaussian_vorticity}
\end{equation}
 $\omega_{v}$ being the peak vorticity. The angular velocity of a
particle around the origin is therefore given by 
\begin{equation}
\Omega(r)=\int_{0}^{r}2\pi r\omega dr=\frac{\pi\omega_{v}r_{v}^{2}}{2}\left(1-e^{-r^{2}/r_{v}^{2}}\right)=\frac{\Gamma}{2}\left(1-e^{-r^{2}/r_{v}^{2}}\right).\label{eq:particle_omg_gaussian}
\end{equation}

The $\Gamma\left(=\pi\omega_{v}r_{v}^{2}\right)$ in equation \ref{eq:particle_omg_gaussian}
is strength of the Gaussian vortex. Figures \ref{fig:Caustics-Gaussian}
show that the vortex strength plays much the same part for Gaussian
vortices as it does for point vortices. The Stokes number of a particle
may then be defined using the characteristic radius of the vortex:

\begin{equation}
St=\frac{\Gamma\tau}{r_{v}^{2}}\label{eq:particle_Stokes}
\end{equation}

Figures \ref{fig:Caustics-Gaussian} plot the analogues of figure
\ref{fig:Caustics_combined}. The actual value of $r_{v}$ does not
make a difference to critical radius $r_{cr}$. However, the nondimensional
parameter $\omega_{v}\tau$ (which is a multiple of the particle Stokes
number and hence depends on $r_{v}$) limits the occurrence of caustics.
No caustics occur for $\omega_{v}\tau\apprle4$. Therefore, as the
width $r_{v}$ of a Gaussian vortex increases in time due to viscosity,
only particles of larger inertia can undergo caustics at later times.

\section{Several vortices}

\label{sec:five}
\begin{figure}
\includegraphics[width=1.0\columnwidth]{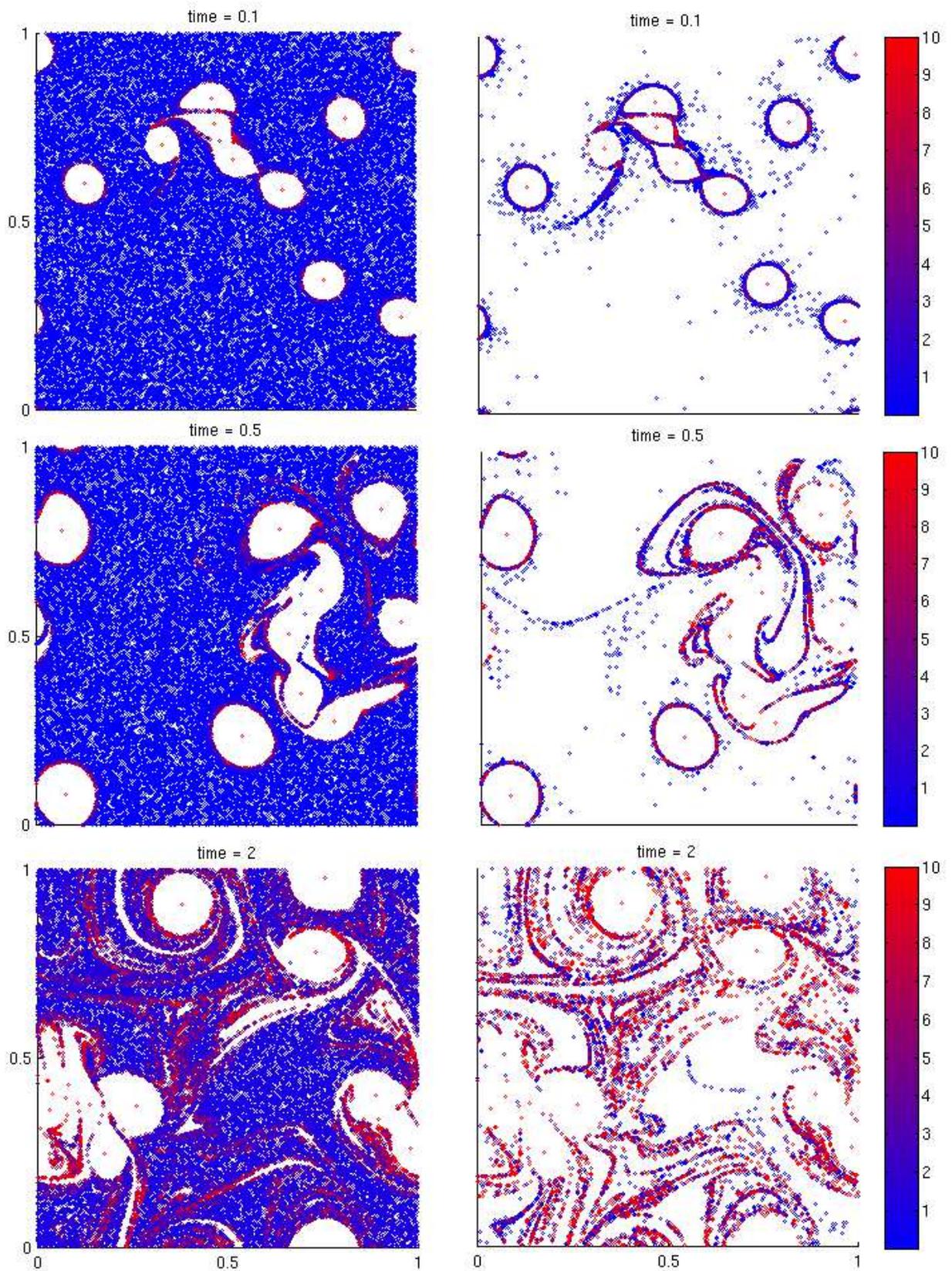}
\caption{\label{fig:10vor snapshots}Snapshots at time $t=200\tau$ of particle
positions with 10 vortices in a doubly periodic $10\times10$ box.
The colour indicates local particle density. The particles have $\tau=0.001$.
Left: Caustics excluded; right: Only caustics.}
\end{figure}

\ADD{We now apply the same density tracking procedure as in sections
\ref{sec:three} and \ref{sec:four} to a system of particles in the
flow of several point vortices in a periodic box. The system of point
vortices is for us a proxy for two-dimensional turbulence. (While
Gaussian vortices would have been more realistic, no expression exists of
the form \ref{eq:Hamiltonian} for Gaussian vortices in a periodic
box.) } \REM{To get some idea of how well our results hold for two-dimensional
turbulence,}

We perform simulations with $N=10$ point vortices of the same circulation
but random signs (zero net circulation) in a doubly periodic box.
The motion of the vortices is computed using the Hamiltonian for this
problem \cite{O'Neil}: 
\begin{equation}
\mathcal{H}=-\frac{1}{2\pi}\sum_{j<k}\Gamma_{j}\Gamma_{k}\left[\mbox{ln}\left|\vartheta\left(\frac{z_{jk}}{L_{x}}\right)\right|-\frac{\pi}{\Delta}\left(\mbox{Im}\left(z_{jk}\right)\right)^{2}\right]\mbox{, }\label{eq:Hamiltonian}
\end{equation}
 where $z_{j}$ is the position of the $j$th vortex in the complex
plane: $z_{j}=x_{j}+iy_{j}$; $L_{x}$ is the length of the box along
the x-direction and $\Delta=L_{x}L_{y}$ is the area of the box.

\begin{figure}
\includegraphics[width=1.0\columnwidth]{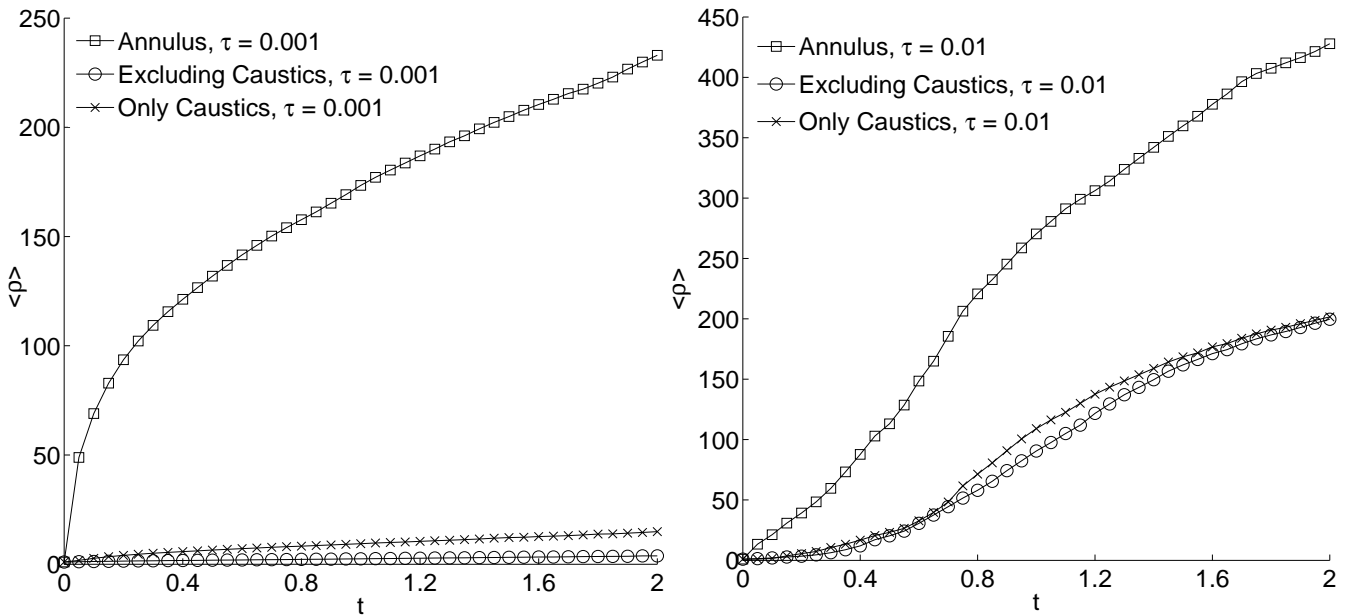}
\caption{\label{fig:10vor density}Average Lagrangian densities. The three
curves in each plot correspond to particles starting well within,
well outside, and in the annulus around the critical radius $r_{cr}.$
Particles starting close to $r_{cr}$ are the primary contributors
of high density regions. Left: $\tau=0.001$, Right: $\tau=0.01$}
\end{figure}

\begin{figure}
\includegraphics[width=1.0\columnwidth]{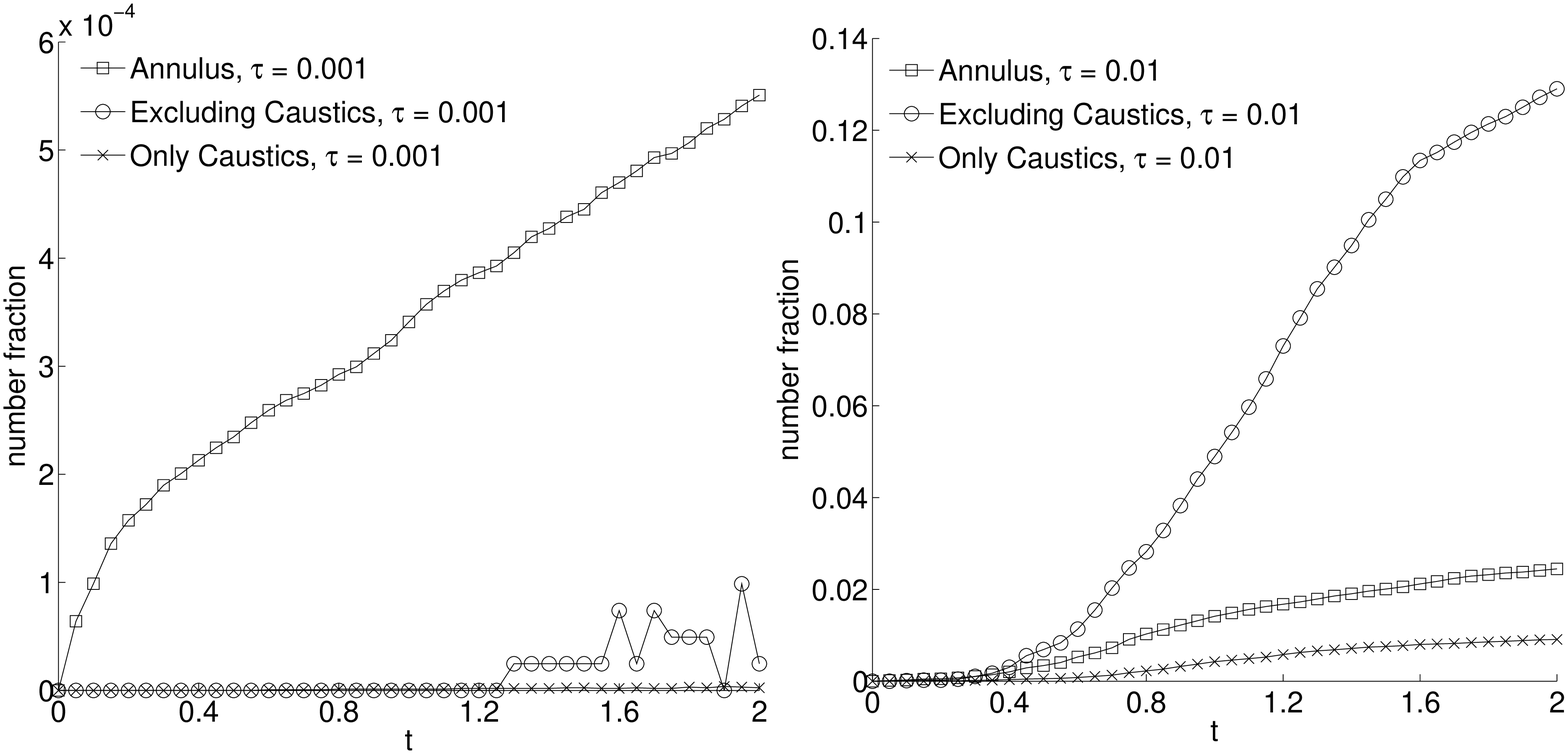}
\caption{\label{fig:10vor tail}Fraction of particles in high-density regions.
We plot the fraction of particles whose neighbourhoods have density
greater than a certain threshold, here prescribed to be $\rho_{cutoff}=1000\rho_{0}$,
where $\rho_{0}$ is the uniform initial density. Left: $\tau=0.001$,
Right: $\tau=0.01$. }
\end{figure}

We study statistics of the particle density as a way of quantifying
the effects of caustics in such systems. We perform simulations of
three types: (i) by distributing particles at random but removing
all particles inside the caustic-producing region of each vortex.
In simulations of type (ii) we do the reverse, i.e., only keep particles
within the critical $r_{cr}$ for caustics formation. In simulations
of type (iii), we start particles in the annular region $0.8r_{cr}\leq r\leq1.2r_{cr}$.
Simulations excluding caustics are done with $4\times10^{4}$ particles;
and simulations with particles only in the caustic regions, and particles
in an annulus are done with $10^{4}$ particles each. We hold $\tau$
constant during each simulation. Figure \ref{fig:10vor snapshots}
shows particle density in the domain at different times, for initial
particle locations exclusively within, and strictly outside, the caustics-forming
region, i.e., for particle distributions corresponding to (i) and
(ii) above. We notice that the highest densities occur in both cases
in the same regions. These appear to correspond to initial conditions
at the edge of the caustics-forming region, i.e., around the critical
radius. To confirm this, we perform simulations of type (iii). Figures
\ref{fig:10vor density} show how significant the region around the
critical radius is for producing high density particles, in comparison
to the regions well inside and well outside. We also note that the
effects of the critical annulus region are more pronounced for the
smaller of the two inertia values.

\ADD{Another point of interest is that there appear to be, in figure \ref{fig:10vor snapshots}, "voids" in the distribution of particles in space, with the radius of the voids given by the critical radius. It is known from simulations of two- and three-dimensional turbulence \cite{Bofetta_etal_2004, Bec_etal_2007} that such voids exist. It is also known that, for small inertia, the voids increase in size with increasing inertia. It is subject to future confirmation whether our idea of a critical radius $r_{cr}$ for inertial particle caustics is directly connected to these voids (and the dependence of their size on particles inertia). The fact that caustics formation is much faster than flow time-scales, and that a given particle inertia leads to a characteristic void size lends credence to this.}

The number of particles with greater than a threshold density, as
figure \ref{fig:10vor tail} shows (we have checked that the exact
value of the threshold does not matter; for this figure a value of
$1000$ was used) is also much higher for particles starting in the
annulus. In fact for the smaller of the two inertias, almost all `caustic
events' happen for particles starting in the annulus.

\begin{figure}
\includegraphics[width=0.5\columnwidth]{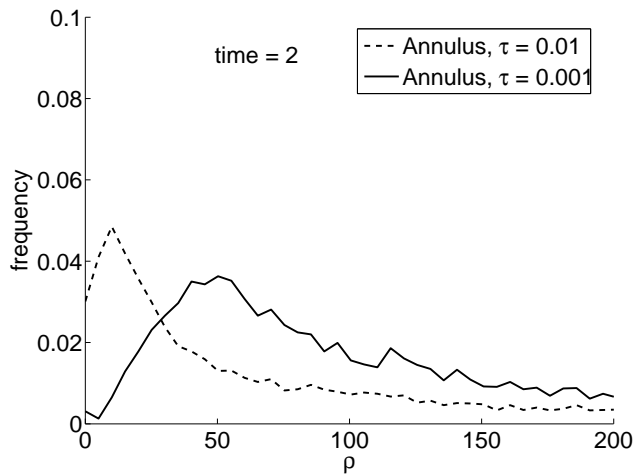}
\caption{\label{fig:10vor_annulus_hist}The density distribution for particles
starting in the annulus regions for two different values for inertia.
The curves for the larger inertia seem to shift towards lower density
values. \REM{Cf. figure \ref{fig:10vor hist}.}}
\end{figure}

We plot the distribution of Lagrangian densities for the many-vortex
simulations. The plots for particles starting in the 'only caustics'
and 'excluding caustics' corroborate the results from section \ref{sec:three},
figure \ref{fig:Density}: particles starting outside the caustics
region show a monotonically increasing density in their vicinity,
whereas particles starting close to vortices get thrown out violently
and so a peak in the distribution is seen at very small density. In
figure \ref{fig:10vor_annulus_hist}, we plot the density distributions
for particles starting in the annulus regions for two different values
of particle inertia. The curves are plotted at the same advection
time (for identical fluid flows, since the particles don't affect
the flow), which is a different multiple of each different inertia.
The curves show the smaller of the two inertias reaching a higher
average density at the same advection time, which is expected, since
the heavier particles have had less time in units of $\tau$ to cluster.

\section{Conclusion}

\label{sec:six}

To summarise, we have studied the formation of caustics in model two-dimensional
vortical flows. For a single vortex, we show that only particles which
are initially within a critical radius $r_{cr}\sim0.5(\Gamma\tau)^{1/2}$
from its centre can cause caustics to form. For a point vortex, particles
placed inside this critical radius can only overtake rings of neighbouring
particle rings which lie within some initial separation $(\Delta r)_{cr}$.
The time for such overtaking is the caustics formation time $T_{caustics}$.
This quantity diverges for $r_{0}\to r_{cr}$. The initial separation
within which caustics form depend on initial conditions and so does
the formation time.

The formation of caustics is intimately related to the creation of
regions of high particle density. Our adoption of Osiptsov's method
enables us to track densities in the neighbourhood of Lagrangian particles
\REM{with high accuracy}. Interestingly, particles which lie close
to the critical radius, i.e. on the edge of the caustics-producing
region, give rise to the largest spikes in density, suggesting that
this region should contribute to the effects that particle caustics
have on the flow. Indeed, we find that for small inertia, almost all
the instances of very dense clusters come from the annular region.
In these regions particle collision frequency is likely to be the
highest \REM{,and water droplets in such regions in clouds can grow substantially
and could contribute to hastening the formation of rain}. Further studies
are needed to confirm this prediction. Outside the critical radius,
particle density displays a qualitatively different behaviour. Density
increase is gentle and monotonic in this case for all time.

\REM{Mehlig and Wilkinson argue that, for particles of very small
inertia, the effects of the centrifugal force are negligible in causing
the collisions necessary to produce rain; that caustics are necessary
for the activation of rain\cite{M-W 2005EPL}.} Our results above
show \REM{that this is true in the sense} that the centrifugal force
can only create caustics in a small area \ADD{around a vortex}.
We also show that the biggest density spikes happen not for particles
that start well within the caustic regions, but rather for particles
that start on the edge of the caustic regions \ADD{(i.e. for particles
which may be considered to have Stokes numbers of order $1$.} We
find also that the effects of this annular region around the critical
radius are more prominent for the smaller of the two inertias.

\ADD{Our results predict that particle collisions in flows with strong
vortices are more probable at a critical radius from the vortex. Our
future work will test this prediction with more realistic flows than
a system point vortices. Whether the existence of this annular band
depends on the confinement offered by the two-dimensional nature of
our setup is also worth checking.} We hope that our work will motivate
viscous three-dimensional simulations to check our predictions on
the importance of the critical annular region about each vortex in
the creation of the densest clusters. In particular it would be useful
to ask whether the extent of clustering can be directly correlated
to suitably weighted summations of square roots of instantaneous circulations
in the flow.

\bigskip{}

\section*{Acknowledgements}

We thank Raymond Shaw, who suggested we study caustics using our system
of point vortices. We thank Nick Ouelette and Jeremie Bec, discussions with whom
led to the discussion of voids in particle distribution in section \ref{sec:five}. 
We also thank the two referees whose suggestions
led among other things to the detailed physical scaling arguments
in section \ref{sec:three}, and a discussion of the validity of the
Lagrangian density tracking algorithm in the appendix. This work is partially 
supported by the Ministry of Earth Sciences, Government of India, 
under the Monsoon Mission Project on the Bay of Bengal.

\bigskip{}

\appendix
\textbf{Appendix: \ADD{Lagrangian} Density tracking}

The density tracking equations of Healy \& Young \cite{Healy-Young 2005}
are modified for a cylindrical polar coordinate system. We follow
the same notation as them. They are 
\begin{eqnarray}
J_{rr} & = & \frac{\partial r}{\partial r_{0}}\nonumber \\
w_{rr} & = & \frac{\partial J_{rr}}{\partial t_{p}}=\frac{\partial v_{r}}{\partial r_{0}}\nonumber \\
\frac{\partial w_{rr}}{\partial t_{p}} & = & \frac{\partial}{\partial r_{0}}\frac{\partial v_{r}}{\partial t_{p}}=\frac{\partial}{\partial r_{0}}\left(-\frac{v_{r}}{\tau}+r\omega^{2}\right)=-\frac{w_{rr}}{\tau}+J_{rr}\omega^{2}+2\omega rw_{\theta r}\nonumber \\
J_{\theta r} & = & \frac{\partial\theta}{\partial r_{0}}\nonumber \\
w_{\theta r} & = & \frac{\partial\omega}{\partial r_{0}}\nonumber \\
\frac{\partial w_{\theta r}}{\partial t_{p}} & = & \frac{\partial}{\partial r_{0}}\frac{\partial\omega}{\partial t_{p}}=\frac{\partial}{\partial r_{0}}\left(\frac{\Omega-\omega}{\tau}-2\frac{v_{r}\omega}{r}\right)=\frac{1}{\tau}\left(\frac{\partial\Omega}{\partial r}J_{rr}-w_{\theta r}\right)-2\left(\frac{\omega}{r}w_{rr}+\frac{v_{r}}{r}w_{\theta r}-\frac{v_{r}\omega}{r^{2}}J_{rr}\right)\label{eq:Osiptsov_density_tracking}
\end{eqnarray}

The equations \ref{eq:Osiptsov_density_tracking} can be integrated
in time to find the density given by 
\begin{eqnarray}
J & = & \left|rJ_{rr}J_{\theta\theta}-J_{\theta r}J_{r\theta}\right|=\left|rJ_{rr}\right|\nonumber \\
\rho_{p} & = & \frac{\rho_{p,0}r_{0}}{J}.\label{eq:post-processsed density}
\end{eqnarray}
 The first of equations \ref{eq:post-processsed density} includes
the factor $r$ because of the Jacobian involved in transforming from
Cartesian to polar coordinates.

\ADD{\textbf{A comment on the validity of Lagrangian density tracking}
Osiptsov's method as described above calculates the density around
a particle as it moves along on its trajectory, and does this without
reference to how many other particles are around the particle being
tracked. We have found that Lagrangian density tracking is better
than the usual way of calculating density (which is to count the number
of particles per unit area/volume). However, the density predicted
by the method will only be realised in a real flow if there is in
fact a continuum of particles that converge onto (or diverge away
from) the particle being tracked. Since this is hardly observed in
real flows (in clouds, for example, there is on average one particle
per every few Kolmogorov boxes), the answers from Osiptsov's method
must be taken with a pinch of salt.}

\label{References}

\end{document}